\documentclass[amsmath,amssymb,twocolumn]{aastex6}
\usepackage{graphicx}
\usepackage{bm}
\usepackage{calc}
\usepackage{color}
\usepackage{url}
\usepackage{array}
\usepackage{tabularx}

\newif\ifAMStwofonts
\AMStwofontstrue

\def\hvn{\hat {\vr}}

\def\gsim{~\rlap{$>$}{\lower 1.0ex\hbox{$\sim$}}}

\def\simpropto{\lower.2ex\hbox{$\; \buildrel \propto \over \sim \;$}}
\def\ltsim{\lower.5ex\hbox{$\; \buildrel < \over \sim \;$}}
\def\gtsim{\lower.5ex\hbox{$\; \buildrel > \over \sim \;$}}
\def\ltsim{\lower.5ex\hbox{$\; \buildrel < \over \sim \;$}}
\def\gtsim{\lower.5ex\hbox{$\; \buildrel > \over \sim \;$}}

\def\sun{\Theta_{_{\mathrm un}}}

\def\kms{\mbox{km\,s$^{-1}$}}

\def\kms{\ {\mathrm km\,s^{-1}}}

\def\pmb#1{\setbox0=\hbox{#1}%
\kern-.025em\copy0\kern-\wd0
\kern.05em\copy0\kern-\wd0
\kern-.025em\raise.0433em\box0}

\def\vr{\boldsymbol{r}}

\def\simlt{\lower.5ex\hbox{$\; \buildrel < \over \sim \;$}}
\def\simgt{\lower.5ex\hbox{$\; \buildrel > \over \sim \;$}}

\newcommand{\beq}{\begin{equation}}
\newcommand{\eeq}{\end{equation}}
\def\beqa{\begin{eqnarray}}
\def\eeqa{\end{eqnarray}}
\def\fixit#1{}

\def\apjs{ApJ. S}

\def\apjl{ApJL}

\def\aap{A\&A}

\begin{document}
\title{On Testing the  Equivalence Principle with Extragalactic Bursts}

\author{Adi Nusser\altaffilmark{1}}

\affil{Department of Physics and the Asher Space Research Institute, Israel Institute of Technology - Technion, Haifa 32000, Israel}
\altaffiltext{1}{adi@physics.technion.ac.il}
\begin{abstract}
An interesting  test of  Einstein's equivalence principle (EEP)
relies on  the observed lag in  arrival times of photons emitted  from extragalactic transient sources. Attributing the lag between  photons 
of different energies to the gravitational potential of the Milky Way (MW),  several authors derive new constraints on deviations from EEP.
 It is shown here that potential fluctuations from the large scale structure are at least two orders of magnitude larger than the gravitational potential of the MW. 
Combined with the larger distances, for sources at  redshift $z\gtsim 0.5$ the {\it  rms} of  
the contribution from  these fluctuations exceeds the MW by more than  4 orders of magnitude.
 We provide  actual  constraints for several objects based on a statistical calculation of the large scale fluctuations in the standard 
$\Lambda$CDM cosmological model.
\end{abstract}


\maketitle

\section{Introduction}
Any deviation  from EEP will have far reaching consequences on 
all fundamental theories of physics \citep{Will2006}. 
The inability to distinguish between properties of motion in    non-inertial frames of reference and certain gravitational fields in
inertial frames \citep{Landau1975}  implies that  the world line of a massless particle  is independent of its energy.
Delays between arrival times of different types of radiation from astronomical burst events
have been proposed \citep{Krauss1988,Sivaram1999} to constrain deviations from EEP through the effect of Shapiro (gravitational) time delay \citep{Shapiro1964}. 
Recently, Gao et al and  Wei et al  \citep{Gao2015,Wei2015} have applied this test to gamma-ray bursts (GRBs)  and Fast Radio Bursts (FRBs) \citep{Lorimer2007}. 
Their strongest constraints are based on FRBs. 
Photons with different frequencies, $\nu$, from these millisecond transients are observed  to arrive at different times. 
The observed time delay (with respect to a reference frequency) follows  $\Delta t_{\mathrm obs}\sim  \nu^{-2}$ 
as expected from the  dispersion of radio waves propagating in an ionized medium. The dispersion measure (DM) is 
large,  indicating sources of  cosmological origin. 
Constraint on deviations from EEP are obtained   by taking  $\Delta t_{\mathrm  obs}$ as an upper limit on the 
difference between Shapiro time delay for photons at two distinct frequencies. 

Adopting the parametrized  post-Newtonian approximation (PPN), deviations from  EEP
are described in terms of the parameter $\gamma$ \citep{Will2006}, where $\gamma=1$ in general relativity. 
The Shapiro time delay is then 
\begin{equation}
\label{eq:SHAPstat}
 t_{\mathrm  gra} = -\frac{1+ \gamma}{c^3}\int_{r_{\mathrm  o}}^{r_{\mathrm  e}}U(\vr(t),t) dr \; , 
\end{equation}
where  the integration is along the path of the photon emitted at $r_{\mathrm e}$ and received at $r_{\mathrm o}$. 
Gao et al and Wei et al  focus on  the contribution from the gravitational potential of the MW. Assuming a Keplerian 
potential for the MW they use corresponding shift 
\begin{equation}
\label{eq:gMW}
\Delta t^{\mathrm  MW}_{\mathrm  gra}=\Delta \gamma \frac{GM_{\mathrm  MW}}{c^3} \mathrm{ ln} \left(\frac{d}{b}\right)\; ,
\end{equation}
where $\Delta \gamma$ is the difference between the $\gamma$  value for the two photons, $M_{\mathrm MW}$ is the mass of the MW, $d$ is the distance to the source and $b$ is the impact parameter of 
the light path with respect to the Galactic center. For  $M_{\mathrm MW} =6\times 10^{11}M_\odot$, $d=1500\; \mathrm Mpc$ and 
$b=5\; \mathrm kpc$, this equation gives  $\Delta t^{\mathrm MW}_{\mathrm gra}/\Delta \gamma = 3.5 \times 10^7 \;  \mathrm s$. 
One of the objects Wei et al use is  FRB 110220 \citep{Thornton2013}. 
Since the observed time delay depends on frequency as $\nu^{-2}$,  most of the lag is due to dispersion of  
of photons. Using the observed  DM, the inferred redshift for this object is    $z\sim 0.81$ 
 (corresponding to $d=1500\; \mathrm Mpc$). Taking 
 the 1 second observed shift between arrival times of 1.5GHz and 1.2GHz photons  as an upper limit on  $\Delta  t_{\mathrm gra}$, they 
 obtain $\Delta \gamma< 2.5\times 10^{-8}$. Wei et al point out that this is a conservative upper limit since the 1 second 
 time delay should mostly be sure to the dispersion of radio waves.
 
Wei et al argue  that incorporating the gravitational potential from the large scale   ($\gtsim 10 \; \mathrm Mpc$) structure (hereafter, LSS)
 tightens the constraint,  but 
they do not  estimate  this effect.
 The current paper  assesses  the contribution of 
 the LSS potential field and shows that  it should  greatly exceed the local MW contribution.
 In generalizing Eq.~(\ref{eq:SHAPstat}) to cosmology we assume $i)$ distances well within the horizon, $ii)$ 
the mechanism for EEP breaking is decoupled from the cosmological background and is  induced solely  by 
 spatial fluctuations   of the gravitational potential, $U^{^{\mathrm LS}}$, resulting from the LSS distribution of matter, and 
$iii)$  a PPN for the cosmological metric \citep{Futamase1988,Hwang2008} with $\gamma$  appearing in the time and spatial components of the metric as $g_{00}\approx -(1-2\gamma U^{^{\mathrm LS}}/c^2)$ and 
$g_{ij}= a(t)(1+2 \gamma U^{^{\mathrm LS}}/c^2)$ where $U^{_{\mathrm LS}} \ll c^2$  and $a(t) $ is the scale factor of the Universe.
We write the shift in the arrival times of 
photons of two different frequencies  due to the Shapiro effect  as
\begin{equation}
\label{eq:SHAP}
\Delta t_{\mathrm gra} (\hvn) =
\frac{\Delta \gamma}{c^3}\int_{r_{\mathrm o}}^{r_{\mathrm e}}U^{^{{\mathrm LS}}}(r \hvn , z) a(z) dr,
\end{equation}
where
$r_{\mathrm o}$ and $r_{\mathrm e}$ are  now comoving distances  and $a=(1+z)^{-1}$  corresponds  
to a comoving distance $r(z)$, at a cosmological redshift $z$. 
This cosmological  Shapiro shift may acquire  negative as well as positive values since  $U^{^{\mathrm LS}}$ fluctuates around zero. 

\section{Order of magnitude based on the observed  LSS motions}
The expected amplitude of  LSS potential fluctuations can be found from the observed peculiar velocities  (deviations from a pure Hubble flow), $
v_{\mathrm p}$,  of 
galaxies. For a nearly  homogeneous   matter distribution at early cosmic times, 
 linear theory provides the intuitive relation  $v_{\mathrm p}\sim t g$, where $g$ 
is the gravitational force field generated by  mass density fluctuations and $t\sim H_0^{-1}$ 
is  the age of the universe as the only possible time scale. 
Peculiar velocity data yield a bulk peculiar velocity of $v_p\sim 300\kms$ for the
sphere of radius  $R \sim  100\mathrm Mpc$ around us. This corresponds to a  gravitational potential 
  $U^{^{\mathrm LS}}_{100} \sim v_{\mathrm p} R H_0 \approx (5\times 10^{-3}c)^2  \sim 50  U_{_{\mathrm MW}}$ where  $U_{_{\mathrm MW}}$
is the gravitational  potential depth associated with the MW.  Note  that $U_{_{100}}=\mathcal{O}(10^{-5})c^2  $ is  of the same order of magnitude as the potential fluctuations inferred from the temperature fluctuations  in the cosmic microwave background \citep{Bennett1994}. 
Since the gravitational time lag  is proportional to the line of sight integral over the potential, 
 the contribution from LSS  greatly exceeds that of  the MW. It also dominates the delay due to the passage of photons through  individual clusters of
 \citep{Zhang2016}.

\section{Theoretical estimate based  on $\Lambda$CDM}
\begin{figure}
\vskip 0.1cm
\includegraphics[width=0.9\linewidth]{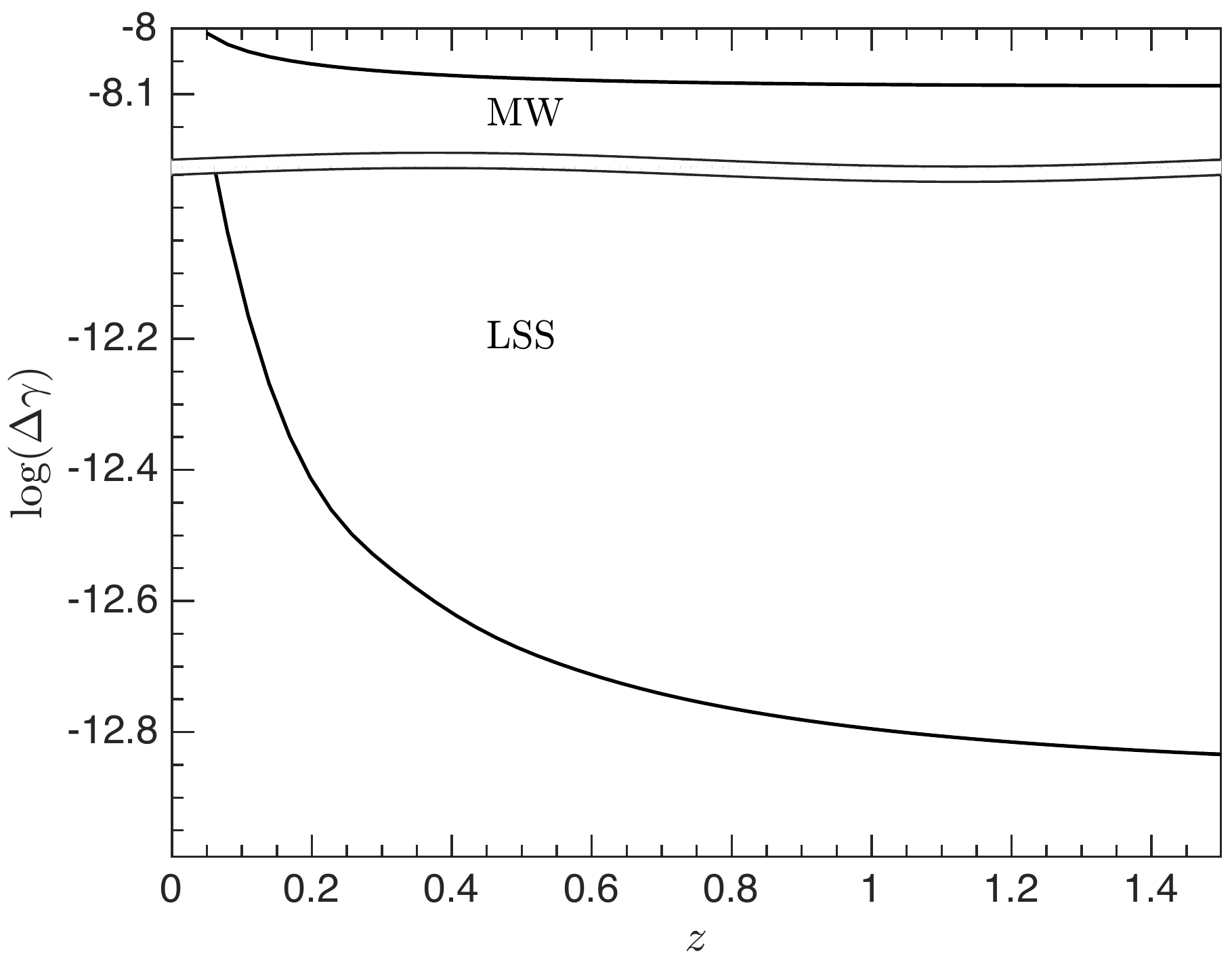}
\hspace*{\fill}
\caption{Upper limits on $\Delta \gamma$ as a function of redshift of the source for an assumed shift of $\Delta t_{\mathrm  gra}<1 \mathrm s$. 
Lower curve corresponds to rms value of the LSS contribution while the upper curve is the limit obtained from the 
MW potential.  These upper limits correspond to a  time lag  of 1 second.}
\label{fig1}
\end{figure}

We provide a statistical estimate of  the shift in the gravitational time lag, $\Delta t^{^{\mathrm LS}}_{\mathrm gra}$, due to     LSS in frame work of the  $\Lambda$CDM 
model \citep{Planck2015}. We are interested in the rms value, $ \sigma= \langle (\Delta t^{^{\mathrm LS}}_{\mathrm gra})^2 \rangle^{1/2}
=\Delta \gamma \tilde \sigma$ where the averaging is over all directions. 
We express 
$
\tilde \sigma^2=\sum_{l}\frac{2l+1}{4\pi} C_l
$ in terms of the angular power spectra, $C_l$,  and write  \citep{Nusser2013}  
\begin{equation}
\label{eq:clphi}
C_{l}  = \frac{2}{\pi c^6}\int dk k^2 P_U(k) \left\vert \int_{r_1}^{r_2} d r D[t(r)] j_l(kr)\right\vert^{2},
\end{equation}
where $P_U$ is the power spectrum of the gravitational potential at redshift $z=0$. Further,  we have used the linear theory 
result that the gravitational potential 
 $U(\vr,t)=(D/a)U_0(\vr,t_0)$  where  $D(t)$ is the linear growth factor \citep{Peeb80}.
Adopting the $\Lambda$CDM cosmology, these expressions are computed numerically as a function of the redshift of the burst. 
The lower curve in  Fig.~\ref{fig1}  is the upper limit $\Delta \gamma < {\tilde  \sigma}^{-1}$ for $\Delta t_{\mathrm gra}< 1 \mathrm  s$ between photons emitted by a burst at redshift $z$. 
For comparison, the   upper curve represents the limit obtained by considering the MW alone according to 
Eq.~(\ref{eq:gMW}). The curve is obtained for 
$M_{\mathrm MW} =2\times 10^{12}M_\odot$  in accordance with the  mass determination from  the dynamics  of the Local Group \citep{Phelps2013a}. 

\section{Actual constraints}
The expected time shift has been estimated in a statistical way. The  {\it rms}  LSS contribution is overwhelmingly 
greater than the MW and, therefore, we could derive  stringe constraints  even without an actual measurement of the   LSS 
gravitational potential to the extragalactic burst. 
 According to the figure, the probability that the LSS contribution at $z\sim 1$  acquires values smaller than the MW's 
 is $10^{-4.8}$, i.e. $4.3\sigma$ rejection level.  Values 200 times smaller than  the MW's are rules out at the $3\sigma $ level. 
 We derive now constraints from several bursts already considered in the literature: 
  \begin{itemize}
\item {\it FRB 110220:} This is the object used in \cite{Wei2015}. Based on the figure,  
 for FRB 110220 at  $z\sim 0.8$ (estimated from the  DM)   
we derive the limit 
\begin{eqnarray}
\nonumber 
\gamma_{_{\mathrm 1.2GHz}} -\gamma_{_{\mathrm 1.5GHz}} &< &4.5 \times 10^{-11} \quad (3\sigma)\\
&< &2.8\times 10^{-12} \quad (2\sigma)\; .
\end{eqnarray}
for $\Delta_{\mathrm gra}<1 \rm s$. 
Since the observed arrival times of photons  depends on frequency as 
expected from the propagation of radio waves in an ionized medium,   we infer  that 
deviations from EEP make a subdominant contribution to the observed lag. This gives us confidence in the 
DM based redshift estimate and  in adopting  the observed lag of 1 second as an upper limit on gravitational delays. 

\item{\it    FRB 150418:   } \cite{Keane2016} associate this FRB with a subsequent fading radio source at the position of a galaxy at 
$z=0.492\pm 008$ (but see \cite{Williams2016} for a different point of view). 
For this redshift,    \cite{Tingay2016}  estimate  $\Delta  t_{\mathrm gra}$  to be less than   5\%-10\%  of the total time delay of 0.8 s.   Following
\cite{Tingay2016}, we take   $z=0.49$ and $\Delta  t_{\mathrm gra}<0.04 \mathrm s$ 
 to obtain the constraint
\begin{eqnarray}
\nonumber 
\gamma_{_{\mathrm 1.2GHz}} -\gamma_{_{\mathrm 1.5GHz}} &< &2.4 \times 10^{-12} \quad (3\sigma)\\
&< &1.4\times 10^{-13} \quad (2\sigma)\; .
\end{eqnarray}
compared to their hard constraint  $\Delta \gamma <10^{-9}$.

\item {\it GRB 090510:}  The firmest constraint obtained in  \citep{Gao2015} 
is for GRB 090510 at $z=0.903\pm 0.003$ \citep{Rau2009} and  a time delay of 0.83 seconds between 
GeV and Mev photons. 
For this object we obtain the limit 
\begin{eqnarray}
\nonumber
\gamma_{_{\mathrm GeV}} -\gamma_{_{\mathrm MeV}} &< &4 \times 10^{-11} \quad (3\sigma) \\
& <& 2.3 \times 10^{-12} \quad (2\sigma)\; . 
\end{eqnarray}

\item{ \it GRB 080319B:}
This GRB is at  $z=0.937$ \citep{Vreeswijk2008} with an upper limit on the time delay of 5 seconds between 
eV and MeV photons. The corresponding limit we derive here is 
\begin{eqnarray}
\nonumber
\gamma_{_{\mathrm eV}} -\gamma_{_{\mathrm MeV}} &<& 2.3 \times 10^{-10} \quad (3\sigma) \\ 
 &<& 1.3 \times 10^{-11} \quad (2\sigma) \; .
\end{eqnarray}

\end{itemize}
These numbers  are smaller  by a factor of a few 100s than the corresponding constraints obtained in \citep{Gao2015}.  
 Future deep galaxy redshift surveys, e.g. Euclid \citep{EuclidRB}, and  peculiar velocity data will
allow an actual estimation of the gravitational potential  along the line of sight to some relevant transient events. 
This should yield robust constraints on the accuracy of EEP from events of cosmological origin. 
Obtaining  redshifts of FRBs is the focus of intense observational activity. Thus  measured redshifts,   especially of repeating events are expected  be available in the very near future. This would be very rewarding since FRBs offer important constraints 
on several aspects of deviations from standard physics
such  the photon mass \citep{Bonetti2016,Wu2016}, in addition to EEP.

\section{Acknowledgments}
The author thanks  Oded Papish, Curtis Saxton, Maciek Bilicki, Enzo Branchini,  Matthias Bartelmann and Yizhong Fan for comments and discussions. 
This research was supported by the I-CORE Program of the Planning and Budgeting Committee, THE ISRAEL SCIENCE
FOUNDATION (grants No. 1829/12 and No. 203/09).
 \bibliography{/Users/adi/Documents/Bibtex/library.bib}
\end{document}

\end{document}